%% file: shai.tex
\documentclass[conference]{IEEEtran}
\usepackage{cite}
\usepackage{amsmath,amssymb,amsfonts}
\usepackage{graphicx}
\usepackage{textcomp}
\usepackage[hyphens]{url}
\usepackage{times}
\usepackage{datetime}
\usepackage{epsfig}
\usepackage[outdir=./]{epstopdf}
\usepackage{endnotes}
\usepackage{graphicx}
\usepackage{color}
\usepackage{amsmath}
\usepackage[T1]{fontenc}
\usepackage{xspace}
\usepackage[normalem]{ulem}
\usepackage{textcomp}

\usepackage{fixltx2e}
\usepackage{booktabs}
\usepackage{caption,setspace}
\usepackage{times}
\usepackage{epsfig}
\usepackage{color}
\usepackage{graphics}
\usepackage{epstopdf}
\usepackage{colortbl}
\usepackage{array}
\usepackage{setspace}
\usepackage{amssymb}
\usepackage{stmaryrd}
\usepackage[labelsep=quad,indention=10pt]{subfig}
\usepackage{algorithm}
\usepackage{algpseudocode}

\newcommand{\sys}{\textsc{\textbf{S}hai}\xspace}
\newcommand{\egsys}{Sys-E}
\newcommand{\thoth}{Thoth\xspace}
\newcommand{\lwcid}{task id\xspace}

\newcommand{\todo}[1]{\textcolor{red}{#1}}

\newcommand{\disj}{\mathrel{\vee}}
\newcommand{\conj}{\mathrel{\wedge}}
\newcommand{\riff}{\mathrel{\text{:-}}}

\newcommand{\says}{\mathrel{\mathsf{says}}}

\newcommand{\pred}[1]{\ensuremath{\mathsf{#1}}}

\def\BibTeX{{\rm B\kern-.05em{\sc i\kern-.025em b}\kern-.08em
    T\kern-.1667em\lower.7ex\hbox{E}\kern-.125emX}}

\begin{document}
\date{}

\title{\sys: Enforcing Data-Specific Policies with Near-Zero Runtime 
Overhead}

\author{
\IEEEauthorblockN{Eslam Elnikety \qquad Deepak Garg \qquad Peter 
Druschel}
\IEEEauthorblockA{Max Planck Institute for Software Systems, Saarland 
Informatics Campus, Germany\\
\{elnikety, dg, druschel\}{{\makeatletter @\makeatother}}mpi-sws.org}
}
\maketitle

\begin{abstract}
\input{abstract}

\end{abstract}

\input{introduction}

\input{overview}

\input{design}

\input{implementation}

\input{evaluation}

\input{related}

\input{discussion}

\bibliographystyle{plain}
\bibliography{shai.bib}

\end{document}

%% file: abstract.tex
Data retrieval systems such as online search engines and online social
networks must comply with the privacy policies of personal and
selectively shared data items, regulatory policies regarding data
retention and censorship, and the provider's own policies regarding
data use. Enforcing these policies is difficult and
error-prone. Systematic techniques to enforce policies are either
limited to type-based policies that apply uniformly to all data of the
same type, or incur significant runtime overhead.

This paper presents \sys, the first system that systematically
enforces data-specific policies with near-zero overhead in the common
case. \sys's key idea is to push as many policy checks as possible to
an offline, ahead-of-time analysis phase, often relying
on \emph{predicted} values of runtime parameters such as the state of
access control lists or connected users' attributes. Runtime
interception is used sparingly, only to verify these predictions and
to make any remaining policy checks. Our prototype implementation
relies on efficient, modern OS primitives for sandboxing and
isolation. We present the design of \sys and quantify its overheads on
an experimental data indexing and search pipeline based on the popular
search engine Apache Lucene.

%% file: introduction.tex
\section{Introduction}
\label{sec:shai_intro}

Data retrieval systems store, aggregate, index, recommend, and serve
information. Examples include large-scale online social media sites,
search engines, and e-commerce sites, but also numerous
organizational, corporate, and government information services. To the
extent that such systems serve personal or private information, they
are subject to various data use policies. 

%Online data retrieval systems such as Facebook, Google, eBay, and 
%Amazon.com serve a corpus of searchable data items. These data items 
%span public documents, web pages, blogs, e-mails, social content, 
%and advertisements.

Each data item served or used by a retrieval system may have its own
\textit{individual} usage policy. For instance, Alice's e-mails are
private to Alice, whereas Bob's e-mails are private to Bob. A
particular blog post of Alice may be public, whereas another of hers
may be available only to her friends. In addition to personal access
control settings, the system may need to to comply with local laws and
regulations regarding data retention and censorship.  Finally, the
system must comply with the provider's own privacy policy, which may
stipulate, for instance, that a user's query and click stream be used
only for personalization.

Ensuring compliance with all applicable data use policies in a large
and agile data retrieval system presents a significant technical
challenge. When compliance checks are entangled with application code,
the policies in effect are difficult to audit and maintain. Moreover,
any application bug or misconfiguration can cause a policy
violation. As a result, there has been significant work on ensuring
policy compliance separate from application
code~\cite{Sen2014sp,thoth}. However, existing systems for compliance
are either limited to type-based policies~\cite{Sen2014sp}, or their
runtime overhead is too high for large-scale data retrieval
systems~\cite{thoth}.

Many of the data use policies that arise in practice are per-user,
data-specific polices. For instance, the EU General Data Protection
Regulation (GDPR) explicitly grants users individual choice regarding
the use of their personal data\cite{gdpr}.  Enforcing individual, 
data-specific
policies, however, requires runtime enforcement for two reasons.
%since individual policies are not amenable to pure static analysis
%techniques for two reasons.
First, static analysis loses precision quickly under data-specific
policies, because the policy in effect for a particular program
variable depends on the value assigned to it at runtime.
%The same front-end program that Alice's friend, say
%Carol, uses to access the (``visible only to Alice's friends'') blog
%post is used by other users who are not necessarily Alice's
%friends. Thus, a given program variable may contain data with very
%different individual policies over time at the same program point and,
%hence, the abstraction of static analysis may lose precision quickly.
Second, policies often refer to information only available at runtime.
For example, whether an access is requested on behalf of user Alice is
known only at runtime. Besides users' identity, data use policies
often refer to users' geographic location (e.g., when a data item is
censored in a specific jurisdiction), wall-clock time (e.g., when a
news ticker item expires at a specific time), or content (e.g., a
user's current list of friends).

\if 0
Therefore, providers interested in enforcing such individual data use 
policies turn to runtime monitoring techniques for policy enforcement, 
such as dynamic information flow control ~\cite{asbestos,flume,thoth}, 
at the expense of incurring runtime overhead. The runtime overhead is 
mainly due to (i) interposing on I/O in order to track taints and to 
evaluate policies, and (ii) recycling processes since Carol's process 
can safely serve another user only after exec'ing. The runtime 
overhead directly increases the providers' operational cost, as they 
either need to sacrifice valuable cycles for runtime policy enforcement 
or need to provision more resources to sustain performance. Hence, 
techniques that reduce the runtime overhead required for policy 
enforcement are critical for adoption in large-scale data retrieval 
systems.
\fi

In this work, we propose \sys, a novel system for policy compliance
that can enforce fine-grained, data- and user-specific declarative
policies with near-zero runtime overhead in the common case.
\sys\ combines \textit{offline, static flow analysis} and \textit{light-weight
  runtime monitoring} using an operating system's capability
sandbox. The idea behind \sys\ is to try and push as much work as
possible to the offline flow analysis to minimize and streamline the
remaining, required runtime monitoring.  The design of \sys\ is based
on the following ideas:

\if 0
%short version
\sys's periodically performs an offline analysis (OA) of the system's
task-level data flows, the policies currently in effect, as well as
other variables observed during prior runs of the system. Based on
this information, the analysis determines the taint acquired by each
of the system's tasks. It then compiles each process's taint into a
set of capabilities for compliant I/O accesses that the process may
perform. These capabilities are then directly enforced by an OS
sandbox that encapsulates each task.  As long as the conditions
anticipated during the static analysis hold at runtime, each task can
perform compliant I/O accesses without runtime intervention, and thus
with minimal overhead.  When the conditions at runtime deviate from
those expected, \sys\ reverts to runtime intervention.
\fi

\paragraph{Use of offline flow analysis}
Many aspects of a data retrieval system's runtime behavior can be
predicted statically and based on runtime monitoring. These aspects
include the normal flow of information among the system's components
(tasks), the set of policies currently in effect, the set of users,
and the geographic region(s) from which a user typically
connects. Based on this information, an offline analysis (OA) predicts
the taint each of the system's tasks will acquire at runtime, subject
to assumptions about the value of runtime variables. Finally, the OA
compiles, for each task and each predicted runtime value, the
predicted taint into a set of capabilities for all compliant I/O
accesses.

\paragraph{Session-level binding of runtime information}
Many runtime variables that are unknown during an offline analysis
become known at the start of a user session. These variables include
the identity of the user, the geographic region from which the user
connects, and the wall-clock time.
%
%, and the exact policies in effect.
%
Based on the actual values of these variables, \sys\ assigns the
appropriate capability set provided by the OA for each task
involved. If the value of a runtime variable is not among those
predicted during the OA, \sys\ registers the value as one that should
be considered during the next OA.

\paragraph{OS sandbox to allow compliant I/O without runtime intervention}
In \sys, each of the system's tasks is encapsulated in an OS sandbox
subject to capability-based I/O access control. When a user session
starts, \sys\ grants each task the capability set predicted by the OA
and selected based on available runtime information. As a result, the
system can perform compliant accesses without runtime intervention.
Because the capability checks are light-weight and performed by the OS
kernel, their overhead is very low. In the common case where the
runtime values are among those predicted by the OA, the cost of
enforcing compliance is near-zero.

\paragraph{Runtime reference monitor as a fall-back}
If a task performs an I/O access for which it does not have a valid
capability, control is transfered to the \sys\ reference monitor (RM),
which performs a runtime policy check. The cause of this event may be
a non-compliant access, an imprecise OA, or a change in the system
state since the OA was performed (e.g., a change in policy, or an
access to content that was created after the OA).  If the access is
non-compliant, the RM denies the offending access. If the access
turns out to be compliant, the RM allows the access and patches the
task's capability set to reflect the latest set of compliant
accesses.

\paragraph{Use of efficient OS isolation primitives}
\sys\ uses light-weight contexts (\textit{lwC}s)~\cite{lwc}, an 
efficient OS
isolation primitive, to isolate multiple user sessions within the same
process, and to isolate \sys's reference monitor.

The rest of this paper is organized as follows. We provide an overview
of \sys\ and its components in Section~\ref{sec:shai_overview} and a
detailed description of \sys's design in Section~\ref{sec:design}. The
\sys\ prototype implementation on FreeBSD is described in
Section~\ref{sec:prototype} and the results of an experimental evaluation
are presented in Section~\ref{sec:eval}. We present related work in
Section~\ref{sec:related} and we conclude in Section~\ref{sec:sum}.

%% file: overview.tex
\section{\sys overview}
\label{sec:shai_overview}

\sys\ is a policy compliance system that helps data retrieval system
providers enforce confidentiality and integrity policies on the data
they collect and serve. We next describe \sys's data flow model,
policy model, overall architecture, and threat model.

\paragraph{Data flow model}

\begin{figure}[tb] \centering
\includegraphics[width=8cm]{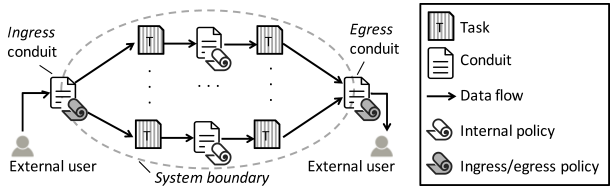}
\caption{\sys\ data flow model.}
\label{fig:dflow}
\end{figure}

Figure~\ref{fig:dflow} shows \sys's data flow model. \sys enforces
data policies in systems that are structured as pipelines of
\emph{tasks}. Each task in the pipeline consumes some data, processes
it, and produces more data, which is then consumed by the next task in
the pipeline. Data enters the pipeline, travels from one task to the
next, and eventually leaves the pipeline in \emph{conduits}, which is
\sys's generic abstraction for any container of data. Files, named
pipes, network connections and tuples in key-value stores are all
conduits. Every conduit has a unique identifier---the full path name
for a file or a named pipe, the five tuple
$\langle$srcIP, srcPort, protocol, destIP, destPort$\rangle$ for a network
connection and the key for an entry in a key-value store. We
distinguish three kinds of conduits: \emph{ingress conduits} that feed
outside data to the initial tasks of the pipeline, \emph{internal
  conduits} that are used to pass data between tasks of the pipeline,
and \emph{egress conduits} that are used to transmit final outputs of
the pipeline to external applications or externally connected users.

\paragraph{Policy model}
An administrator may associate a policy with any ingress conduit. This
policy is a \textit{one-point description} of all the confidentiality 
and integrity requirements of the data entering the system through that
conduit. For example, the policy might say that the data (in the
conduit) should be accessible only to Alice, or that it should be
accessible only to Alice and her friends, or that it should not be
accessible to users connecting from a specific geographic region
(where the content may have been black-listed by legal mandate), or
that the data in the conduit must have a specific shape or type. \sys
also allows policies to be associated with internal conduits, but
these internal policies do not have to be trusted for enforcing the
policies of ingress conduits.

\sys always enforces every conduit's policy, even on data derived
downstream from that conduit's data. For example, suppose an ingress
conduit has a policy ``private to Alice only'' and that the rest of
the pipeline works as follows: The ingress conduit is read by task A,
which writes its output to a file f; file f is read by task B, which
then sends a message to a different user Bob. The last message from 
task B to Bob potentially violates the ingress conduit's policy since it
completes a flow from Alice's private data to Bob, so \sys will not
allow this last message to be sent.

\sys's actual policies can be much richer than those in this simple
example and can specify declassification (i.e., policy relaxation)
based on clock time and the type and content of data. More precisely,
\sys's policies are specified in a declarative policy language, nearly
identical to that used in Thoth~\cite{thoth}. In this language, a
conduit's policy has three rules: 1) A \textbf{read} rule specifies
who can read the conduit's data directly; 2) A \textbf{declassify}
rule specifies what \textbf{read} rules should apply to conduits
downstream in the pipeline, thus controlling who can read derived
data. The \textbf{declassify} rule specifies a set of tests (called
\emph{declassification conditions}) on the global state and data in
any conduit downstream, and how the \textbf{read} rule can be relaxed
when each of those tests is satisfied; 3) An \textbf{update} rule
specifies what type of content can be written to the conduit and by
whom. 

Note that \sys's policies are \emph{data-specific}: Every piece of
data can have its own policy. In particular, two pieces of data of the
same type can have different policies. For example, although a file 
containing Alice's email and a file containing Bob's email have the 
same type ``email'', both have different policies---the former is 
accessible only to Alice while the latter is accessible only to Bob. 
This contrasts with other work like Grok~\cite{Sen2014sp}, which 
enforces only \emph{type-specific} policies, where all data of the same 
type has the same policy.

For brevity, we do not describe the details of the policy
language here. All our development can be followed without
understanding the syntax of the language.

\subsection{Standard solutions and their shortcomings}
At an abstract level, enforcing {\sys}'s policies requires determining,
for each egress conduit, which ingress conduits' data could flow to it,
and what declassification conditions (if any) must be satisfied along 
the flow. This is a standard data flow analysis problem, for which many
different techniques have been proposed in literature. We briefly
outline these existing techniques and their shortcomings in the
context of (\sys-like) data-specific policies.

\emph{Static} techniques determine flows by analyzing the source code
of the system. In addition to requiring the source code and being
language-specific, static techniques work well only when enforcement
is limited to type-specific policies. Static techniques do not work
well when policies are data-specific. The reason is simple: Static
techniques approximate data with program variables; as a result, the
analysis cannot distinguish the policies of different data after they
flow through the same variable. Data-specific policies care about this
difference, while type-specific policies do not.

\emph{Dynamic fine-grained} techniques, also known as runtime taint
tracking techniques, track data flows between program variables, or
between memory objects and machine registers at runtime. Depending on
the specific implementation, a dynamic fine-grained technique may not
have the shortcomings of static techniques mentioned above, but
dynamic fine-grained techniques must intercept all memory and register
reads and writes. This interception makes their overhead prohibitively 
high for most online systems (in the orders of upper 10s to 100s of 
percent).

\emph{Dynamic coarse-grained} techniques track flows at coarser
granularity, typically only across tasks in a system but not within
each task. They only intercept reads and writes to conduits shared
between tasks. This is far more efficient than tracking all reads and
writes to registers and memory. Theoretically, this comes at the cost
of precision---if a task reads a conduit f and later writes a conduit
g, a coarse-grained technique must conservatively assume that there is
a flow from f to g, even if the data written to g was independent of
the data read from f. This can cause overtainting. Practical
experience suggests that in data retrieval systems structured as
pipelines of tasks, this kind of precision loss can be mitigated by a
slight relaxation of policies~\cite{thoth}. Consequently, dynamic
coarse-grained tracking is a reasonable option for enforcing
{\sys}-like policies.

Nonetheless, dynamic coarse-grained tracking \emph{still has a
  significant performance impact}, at least on moderate or high
throughput systems. As a case in point, the Thoth system~\cite{thoth},
which uses dynamic coarse-grained tracking to enforce policies
identical to those considered in this paper, has a relative overhead
of almost 3.5\% on the throughput of a simple data indexing and search
pipeline, even at a very modest throughput of only \textasciitilde 300
queries/s/machine. As the throughput increases, this relative overhead
increases significantly, reaching over 23\% at
\textasciitilde 3,000 queries/s/machine on a port of Thoth to our
experimental setup (see Section~\ref{sec:eval}).

Hence, no existing system can enforce data-specific policies with
consistently low overhead. Our goal with \sys is to change this state
of the art. Ideally, we want to enforce data-specific policies with
\emph{zero overhead}. Of course, attaining this ideal goal is
impossible but as we show, we get very close.

\paragraph{Thoth}
The starting point for our design is Thoth, which can already enforce
data-specific policies efficiently in small-scale, low throughput
systems. In the following, we explain briefly how Thoth works, what
the dominant sources of overhead in Thoth are, and what {\sys} does
differently to mitigate these overheads.

As stated above, Thoth performs coarse-grained runtime flow tracking
to enforce policies. Thoth maps tasks to OS processes and implements a
reference monitor (RM) that intercepts every conduit I/O in the
kernel. The RM maintains a \emph{taint} for every task (process) in
the pipeline. This taint is actually a policy that is always at least
as restrictive as the policies of all conduits that the task has
read in the past.

When a task opens a conduit for reading, the RM intercepts to check
whether the taint on the task is already more restrictive than the
policy of the conduit. If so, it does nothing further. If not, it
intersects the current taint of the task with the policy of the
conduit. When a task opens a conduit for writing, the RM intercepts
to first check the \textbf{update} rule of the conduit. Next, it
checks the declassification conditions in the taint of the task,
which may relax the taint, and then checks whether the (possibly
relaxed) taint is at least as permissive as the policy of the conduit
being written. These checks on conduit opens ensure that, modulo
declassification, the policies of conduits downstream of a conduit f are
always more restrictive than f's policy. As a result, the restrictions
of f's policy cannot be ``lost'' on data derived downstream.

The RM \emph{enforces} policies on egress conduits connected to
end-users by direct checks. For example, if an egress conduit's
\textbf{read} policy says that only Alice can read, then the RM
ensures that the egress conduit is actually connected to Alice by
verifying the public key that authenticates the connection.

As mentioned above, despite its efficiency compared to older
solutions, Thoth still has significant overhead with respect to the
system's throughput. This overhead has two dominant sources.
\begin{enumerate}
\item Interception of every conduit open by the RM to check taints and
  declassification conditions is expensive. In Thoth, every
  interception involves a context switch to a dedicated process that
  hosts the RM.
\item Once a user-facing task has served private data to a user, that
  task cannot serve a different user without shedding its previous
  taint. To shed that taint cleanly, the task must be reset to a clean
  state. The usual way of doing this is to re-exec the process hosting
  the task. Re-execing is expensive. Since taint must be shed only
  once per user session, the amortized cost of re-execing reduces with
  increase in session length, but it is still significant even for
  moderate session lengths (4-8 user queries per session) in Thoth.
\end{enumerate}

\subsection{\sys: Key ideas}

\sys is a \emph{re-design} of Thoth with two key ideas to mitigate
most of Thoth's overhead. First, {\sys} adds to Thoth a new offline
phase that does most of the work of the RM ahead-of-time, thus
significantly reducing the need to intercept I/O. Second, \sys uses a
different implementation of tasks that allows for much faster state
reset. End-to-end, the offline phase reduces the overhead on each user
request to near-zero in the common case, while the change to the
implementation of tasks significantly reduces the overhead on user
session establishment.

\paragraph{Eliminating RM interceptions}
\sys eliminates the need for runtime interception of most conduit
accesses using a periodic, ahead-of-time, offline analysis
(OA). During the OA, \sys makes (and caches) policy checks on the
reads and writes that the system is likely to make in later executions
of the pipeline. For this, the OA takes as input a list of tasks in
the pipeline, what conduits each task is likely to read and write
during the pipeline's execution, an estimate of the task's anticipated
taint at runtime and the policies of all conduits. With the exception
of the policies, these inputs are \emph{not trusted} for policy
enforcement; getting some of them wrong only results in a
proportionally higher overhead at runtime. All inputs can be easily
determined by running the pipeline in a test environment, by
monitoring the production system, or by a simple manual analysis.

The OA simulates the checks that the Thoth RM would make for each
conduit access specified in the inputs (but without actually running
the pipeline). Later, at runtime, each task runs in a \emph{OS
  sandbox}, which allows conduit accesses that were already checked by
the OA without faulting into the RM. These accesses run at native
speed. In the rare case that an access not foreseen by the OA occurs,
the OS sandbox faults into the RM, which makes the same policy
checks that Thoth would make.

Our current prototype uses FreeBSD's capability system
(Capsicum)~\cite{capsicum} (with small modifications) for the OS
sandbox. Capability checks in Capsicum are highly optimized and incur
nearly zero overhead. This, coupled with the OA, reduces the overhead
of I/O interception to nearly zero in the common case.

While this idea is conceptually simple, it has several nuanced details
that we explain in Section~\ref{sec:design}. First, a task's ability
to make certain accesses may depend on parameters whose exact values
will be known only at runtime, e.g., which user has authenticated
remotely, which geographic region the user has connected from (to
enforce legal, region-based content blacklisting), etc. To permit the
OA to take these parameters into account, the anticipated values of
these parameters can be coded within the task description
(specifically, within the task's taint). These anticipated values are
then verified at runtime, but only \emph{once} when the task starts
running. In practice, this amounts to checking these parameters once
per user session, not on every conduit access, which makes the checks
efficient.

Second, a task's ability to make accesses may depend on meta-data such
as friends lists being in a certain state (e.g., Alice can access
Bob's friends-only content only while she is Bob's friend). This state
may change after the OA has finished, thus (partially) invalidating the 
OA's analysis. Consequently, the OA must inform the RM of such meta-data
dependencies and the RM must track runtime updates to meta-data
occurring in the dependencies to avoid policy violations.

\paragraph{Reducing the cost of task reset}
The need to reset a user-facing task between sessions of two different
users is fundamental to coarse-grained taint tracking and cannot be
eliminated entirely. To reduce the cost of this reset significantly,
\sys relies on a recent OS primitive called light-weight contexts
(\textit{lwC}s)~\cite{lwc} to rollback the state of a user-facing task
to a clean state efficiently. \textit{lwC}s support multiple tasks
with separate address spaces and file descriptor tables \emph{within}
the same process. Resetting an \textit{lwC}'s state resets only the
``essential'' elements (the memory mappings and open file descriptors)
and is faster than re-execing an entire process. This cuts down
overheads significantly compared to Thoth.

As an added benefit, the use of \textit{lwC}s also allows implementing
the RM itself in a \textit{lwC}, in place of a separate process (as in
Thoth). This reduces the cost of interception for the few reads/writes
that fault into the RM in \sys from a standard OS context switch to a
\textit{lwC} switch, which is cheaper since it does not involve
scheduling delays. We describe \textit{lwC}s and their use in \sys in
Section~\ref{sec:prototype}.

\subsection{Threat model}

Like Thoth and almost all other work on information flow control, the
goal of \sys is to ensure that policies on ingress conduits are
enforced despite bugs in the system's implementation. The concern is
inadvertent data leaks, not extraction or stealing of information by
malicious adversaries. As such, low-level vulnerabilities (buffer
overflows, control flow hijacks, etc.) are not a concern. Implicit
flows and side-channels like timing channels would, in principle, be a
concern in this setting, but \sys focuses only on the larger, more
prominent risks from explicit leaks of data.

Since \sys is primarily a userspace system, the kernel (including its
sandboxing mechanism) is trusted. \sys's integral components---the RM
and the OA---are both trusted. Policies on ingress nodes are assumed
to represent privacy requirements correctly and all meta-data (e.g.,
friends lists) on which their interpretation depends is assumed to be
accurate.

Policies of internal conduits can be chosen arbitrarily. Getting these
wrong can block legitimate data flows in the pipeline, but cannot
violate policies of ingress conduits.  Any input provided to the OA,
with the exception of the policies of conduits, is not
trusted. Getting these inputs wrong can only impact performance and/or
functionality, not policy enforcement. However, policies of conduits
provided to the OA must be the same as those used by the RM.

%% file: design.tex
\section{\sys design}
\label{sec:design}

\sys's design consists primarily of two components---the offline
analysis (OA) and a runtime sandbox and reference monitor (RM). We use
Thoth's policy language (with very minor extensions) for representing
policies. In the following, we first present a running example that we
use to illustrate various concepts and that also forms the basis of
our evaluation. We then describe the OA and the runtime system.

%% Our extensions to Thoth's policy language are described in
%% Section~\ref{sec:prototype}.

%% and our
%% modifications to Thoth's policy language.

\subsection{Example: Search pipeline}
\label{sec:design:example}

Our running example, called {\egsys}, is the same as that used in
Thoth's evaluation. {\egsys} models the search component of a typical
user data-driven system such as a modern social platform.

{\egsys} indexes a corpus of heterogeneous data consisting of public
documents (modeling public content on the WWW), documents private to
individual users (modeling content such as emails and individual
calendars), and semi-private documents shared among stipulated subsets
of users (modeling content such as social media posts that are
accessible only to friends or friends of friends). Each piece of
content is stored in a separate file. These files are the ingress
conduits of {\egsys}. The system supports friends lists of users,
which are used by the policy enforcement mechanism. The system also
has lists of blacklisted documents which should not be visible to
users connecting from specific geographic regions to support legal
blocking of content.

%% Friends lists and content blacklists can be
%% updated through separate pipelines whose details are irrelevant to
%% this example.

The first task in {\egsys}'s pipeline is a \emph{data indexer} that
builds an index mapping keywords to documents containing those
keywords. This task consumes all the content files above and produces
the index, which is also stored in files. Note that the data indexer
is mostly offline; it only runs periodically.

The next task in the pipeline is a \emph{search engine}, which accepts
a user query (a set of keywords) over a pipe from a user-facing
front-end task (described next), looks up the index, and responds back
to the front-end with a \emph{list} of documents that contain those
keywords. Technically, the search engine does not return a list of
documents but instead passes open file descriptors for the matching
documents over a pipe.

The last part of our pipeline is the \emph{front-end}, which hosts a
web server through which remote users interact with {\egsys}. For
every incoming user connection, {\egsys} spawns a new
\emph{user-specific worker task}, which authenticates the user (with
the user's public key), and then accepts search queries from the
user. It forwards each search query to the search engine, then reads
each of the matching documents returned by the search engine to
extract a snippet, composes all the snippets into a set of
``results'', personalizes the results using stored preferences of the
connected user, and inserts advertisements to generate revenue. It
then returns the resulting page to the user. Note that this last part
of the pipeline is a not a single task, but consists of a separate
task for every connected user.

%% The relevant privacy goal here is to ensure that the privacy policies
%% of each indexed file are respected. For example, content from Alice's
%% private file should not be sent to Bob even if the indexer, the
%% searcher and the worker task that handle's Bob session are buggy. We
%% describe these policies next.

\paragraph*{Ingress conduit policies}
The \textbf{read} rules of the ingress conduits specify expected
confidentiality requirements: Public documents have an all permissive
read rule (anyone can read them), Alice's private files have a read
rule that allows access to Alice only, and Alice's semi-private files
have a read rule that allows access only to Alice and her friends (or 
friends of friends).

Declassification rules are more interesting. Note that the indexer
consumes the private content of all users and, hence, in principle,
its output (the index) should not be accessible to any user. Since the
search engine consumes the index, its output (file descriptors of
documents that match a user query) must also not be accessible to any
user. This effectively means that the end-user will not be able to see
any output from the pipeline!

To work around this, we relax the policies of all indexed
files. Specifically, the \textbf{declassify} rules of all indexed
files allow a complete declassification into any conduit that can only
transfer open file descriptors but no other content. The pipe from the
search engine to a worker task is such a conduit. This allows the
search engine to return open file descriptors of matching documents to
worker tasks, and allows the pipeline to work as expected. (An
additional check in the kernel, described in
Section~\ref{sec:prototype}, ensures that the worker task can only
receive descriptors that it could have opened itself; this prevents a
buggy search engine from sending a descriptor for Alice's private file to
Bob's worker task.)

\if 0
\footnote{Thoth uses a slightly different declassification to
  the same effect: There, the search engine returns an ascii list of
  document ids that match a search query and the \textbf{declassify}
  rule on all indexable content allows a declassification into a list
  of valid document ids. Enforcing Thoth's \textbf{declassify} rule
  requires an expensive runtime check on the data sent from the
  search engine to the front-end. In contrast, our modified rule can be
  implemented very cheaply by a small kernel modification, which we
  describe in Section~\ref{sec:prototype}.}
\fi

The declassification policies of indexed content also have additional
clauses for enforcing region-specific censorship. A user's profile 
(including preferences)
has a policy that allows access only to the user. We elide the details
of these policies here. The paper on Thoth describes these policies in
detail.

\input{sflow}

\input{monitoring}

%% file: sflow.tex
\subsection{The offline analysis (OA)}
\label{sec:sflow}

As its name suggests, the OA is an offline process that runs
periodically on the side, not within the actual system pipeline. The
goal of the OA is to check, ahead of time, which conduits each task in
the pipeline can read and write. Accesses that check successfully in
the OA do not have to be intercepted in the pipeline at runtime, which
reduces runtime overhead. To improve efficiency, the OA should be
configured to check as many accesses as possible ahead of time. Of
course, not all accesses can be checked ahead of time; these accesses
are subject to policy checks by the RM as described in
Section~\ref{sec:monitor}. Accesses to conduits that do not exist when
the OA runs, including pipes, fall in this category. In a properly
configured system, these should be the only accesses that are checked
at runtime.

%% The OA is only a (significant) performance optimization.  For policy
%% compliance, the set of accesses checked by the OA does not have to
%% under- or an over-approximate the actual accesses that will happen at
%% runtime. If the OA does not certify an actual access, that access will
%% be subject to policy checks at runtime by the RM, as described in
%% Section~\ref{sec:monitor}. This has an overhead on the runtime, but is
%% still safe. (Reads and writes to conduits that do not exist when the
%% OA runs, including pipes, fall in this category.)

%% If the OA ends up certifying more accesses than happen at runtime,
%% this only results in some extra work in the OA. In general, for
%% runtime efficiency, the OA should be run to err on the side of
%% checking an over-approximation of the reads and writes of each task.

\paragraph{OA's inputs}
The OA takes the following parameters as inputs:
\begin{enumerate}
\item A list of tasks on which to run the OA. If a task's accesses
  depend on runtime parameters such as the identity of the user the
  task will serve, a separate instance of the task should be listed
  for every combination of these parameters.\footnote{The identity of
    the user is not the only possible policy-relevant runtime
    parameter although, for simplicity, we discuss only this parameter
    here. Another parameter that our implementation of {\egsys} uses
    is the geographic region from which the user connects; we use this
    parameter to enforce region-specific legal blacklisting of
    content.}
\item For each task, lists of conduits whose reads and writes by this
  task have to be checked.
\item The steady-state \emph{taint} of each task. This is explained
  below.
\item The policies of all conduits in the system.
\item Any policy-relevant meta-data such as \egsys's friends lists and
  region-specific content blacklists.
\end{enumerate}

The taint of a task is a policy that the RM associates to the task at
runtime. This policy is always at least as restrictive as the policies
of all conduits that the task has read. \sys enforces this policy on
all data that is output by the task and all data that is derived
downstream from this output data. The relevance of the taint is that
it allows a \emph{local} check to determine if it is safe to allow a
task to read a conduit: The read is safe if the task's taint is at
least as restrictive as the conduit's policy since, then, the
conduit's policy is guaranteed to be enforced downstream. Input (3) to
the OA asks for the runtime steady-state taint of each task.

All inputs (1)--(5) can be determined fairly easily. (1) follows from
the schema of the pipeline and, for parametrized tasks, from the
possible values of the parameters (e.g., the list of registered
users).

The lists in (2) should include as many runtime accesses of the task
as possible. These accesses can be determined either by testing,
monitoring the production pipeline or manual analysis. For simple
pipelines, manual analysis may be straightforward. This works, for
instance, for \egsys: The indexer reads all indexable content and
writes the index; the search engine reads the index and indexed content but
writes to a pipe that is created only at runtime, so the write is
irrelevant for the OA; a user's worker task should read only content
that is accessible to the user (the user's own private content,
content shared by her friends with their friends, public content,
etc.) and it writes to a network connection that is also created only
at runtime, so this write is also irrelevant for the OA.

(3) can be determined by simple manual analysis, testing or monitoring
of the production pipeline. For example, in the {\egsys} pipeline,
ignoring region-specific censorship for simplicity, the taints are
fairly straightforward. (a) Indexer and search engine: Disallow any reads,
but eventually allow declassification into a conduit that can only
transfer file descriptors, (b) User X's worker task: Only X can read.

(4) and (5) should be readily available in the system's meta-data.

\sys includes a dedicated language to represent (1)--(5); we elide the
details of the syntax here.

\paragraph{OA's operation}
The OA checks relevant policies for every conduit read and write
mentioned in input (2) and determines which reads and writes are
policy compliant and which are not. For simplicity, we first describe
the checks assuming that there are no declassification conditions in
policies. We then describe the changes needed to handle
declassification conditions.

In the absence of declassification conditions, the checks that the OA
makes are conceptually straightforward. A task T can \emph{read} a
conduit f if f's \textbf{declassify} rule (the rule that governs the
use of f's data downstream) is at least as \emph{permissive} as T's
taint. This ensures that f's data remains protected downstream in
accordance with f's policy. Dually, a task T can \emph{write} a
conduit f if f's \textbf{declassify} rule is at least as
\emph{restrictive} as T's taint. This ensures that T's taint is
respected on all of T's outputs downstream. For a write, the OA
additionally checks that f's \textbf{update} rule is satisfied.
%Additionally, for the write, f's \textbf{update} rule must
%be checked.

When policies have declassification conditions, the check for reads
remains unchanged. However, the policy comparison check for writes is
more elaborate. The OA first checks if any declassification conditions
in T's taint are satisfied. If so, it creates a list of T's updated
taints, with one taint for every satisfied declassification
condition. If not, it creates a list with only T's current taint. The
write is deemed okay if f's \textbf{declassify} rule is more
restrictive than any of the taints in the list just created.

As an example, suppose that the OA wants to validate a write to
conduit f by task T when T's taint is ``only Alice can read until the
clock time exceeds midnight on December 31, 2017'' and f's
\textbf{declassify} rule is all permissive. (T's taint allows a
declassification of Alice's private content at the end of 2017.) In
this case, the declassification condition in T's taint is ``until the
clock time exceeds midnight on December 31, 2017''. So, the OA checks
whether the clock time is past midnight on December 31, 2017. If this
is the case, then T's resulting taint imposes no restrictions, so the
write is okay. If this is not the case, then the write is not okay.

These conceptually straightforward checks are more nuanced when they
involve meta-data that can change over time. Consider the case of
Alice's worker task in {\egsys} reading a document with the policy
``accessible to Bob's friends only''. In this case, the policy check
above will succeed only if Alice is in Bob's friends list. Suppose
that Alice is in Bob's friends list when the OA runs. Now note that,
in the future, the validity of this check is \emph{conditional} on
Alice remaining in Bob's friends list. If Bob unfriends Alice, this
validity is lost.

Consequently, with each access that it successfully validates, the OA
also returns a list of conditions on the system state under which the
access was validated. We call these conditions the \emph{state
  conditions} of the access. One general state condition is that the
policy of the conduit must be what it was when the OA ran. At runtime,
the RM checks these conditions before using the OA's validations as
explained in Section~\ref{sec:monitor}.

Finally, the validity of the accesses of a task may depend on
parameters that can be determined at runtime only. For example, in
\egsys, a worker task serving user X should be able to read only
conduits that X is allowed to read, but the identity X will be known
only at runtime. In the OA, this is handled by executing the analysis
for all possible instances of the parameter (X in this
case). Technically, the OA is given a separate instance of the task
for every possible value of the parameter. Thus, in \egsys, there is
one instance of the worker task for every registered user---there is a
task called ``Alice's worker'', another called ``Bob's worker'',
etc. The specific value of the parameters for a task instance are
coded in the \emph{taint} of the instance. In \egsys, the taint of
Alice's worker is ``Only Alice can read downstream'', while that of
Bob's worker is ``Only Bob can read downstream''. With these precise
taints, the OA validates all accesses for the specific instances of
the task. At \emph{runtime}, the task must register with the RM as the
correct instance, else it won't be allowed to communicate with the
connected user. Thus, safety is always maintained.

\paragraph{Formal description of the OA's algorithm}
Algorithm~\ref{algo:static} summarizes the work of the OA. The
algorithm does exactly what is described above. The function
isAsRestr($r1$, $r2$) checks that policy rule $r1$ is at least as
restrictive as $r2$ and returns a boolean indicating whether this is
the case ($okay$) and, if so, what parts of the system state were
relevant to this determination (the state conditions, $conds$). The
function isAsRestrWithDeclass is similar but it also applies
declassification within $r2$. The function policyEval evaluates a
policy rule.

All these functions are based on similar functions in Thoth. Thoth
uses these functions at runtime, not ahead-of-time; we modified the
functions to track which parts of the system state are relevant to the
result.

The output of the OA is a list of tuples of the forms (read, T, f,
$conds$) and (write, T, f, $conds$) indicating that task T can
respectively read or write conduit f if the state conditions $conds$
hold on the system state.

\begin{algorithm}[t]
\caption{OA's algorithm}
\label{algo:static}
\raggedright
\textbf{Inputs:} (1)--(5) as described in text. In
particular, (2) is a list of expected reads of the form (read, T, f)
and a list of expected writes of the form (write, T, f).

\mbox{}

\textbf{Output:} A list of tuples of the form ([read | write], T, f,
$conds$), meaning that T can read or write f if conditions
$conds$ hold on the system state.

\mbox{}

\begin{algorithmic}[1]
  \State $output$ $\gets$ $\emptyset$
  \ForAll {(read, T, f) in input (2)}
  \State $pol \gets $ f's policy in input (4)
  \State $taint \gets$ T's taint in input (3)
  \State $(okay, conds) \gets$ isAsRestr($taint$, $pol$.\textbf{declassify})
  \If{$okay$}
  \State add (read, T, f, $conds$) to $output$
  \EndIf
  \EndFor
  \ForAll {(write, T, f) in input (2)}
  \State $pol \gets $ f's policy in input (4)
  \State $taint \gets$ T's taint in input (3)
  \State $(okay1, conds1) \gets$ isAsRestrWithDeclass($pol$.\textbf{declassify}, $taint$)
  \State $(okay2, conds2) \gets$ policyEval($pol$.\textbf{update})
  \If{$(okay1$ \&\& $okay2)$}
  \State add (write, T, f, $conds1 \cup conds2$) to $output$
  \EndIf
  \EndFor
  \State \Return{$output$}
\end{algorithmic}
\end{algorithm}

\paragraph{Using the OA in practice}
It may seem that the total work of the OA is enormous: For every task
and every conduit that the task may potentially access, the access
should be validated ahead-of-time by the OA for runtime efficiency. In
the context of \egsys, for example, assuming 10 million users and, on
average, 1,000 pieces of content accessible to each user, this amounts
to 10 billion checks just for the user-specific worker tasks every
time the OA runs. This sounds intractable.

In reality, not all these checks are necessary. We describe two
obvious optimizations. First, the OA's checks only examine the
policies of conduits, not the conduits themselves. Consequently, if a
set of conduits share the same policy, then it is safe to run the OA
on only one of those conduits and transfer the OA's result to all
other conduits in the set. This optimization is quite useful. For
example, all of Alice's private content (like her emails) will have
the same policy. Similarly, all the uncensored public content on the
WWW has the same policy (it is accessible to everyone).

Second, the OA results remain valid until policies or policy-relevant
meta-data change. Consequently, there is no need to include the
content of inactive users in the OA very often. The OA can also be run
on a specific user's content on-demand, e.g., when the system detects
that the user has become sufficiently active.

We quantify the cost of the OA on a realistic but simulated workload
in Section~\ref{sec:eval}.

%% file: monitoring.tex
\subsection{Runtime monitor and OS sandbox}
\label{sec:monitor}

\sys's runtime infrastructure consists of two components. First, we
rely on an existing OS light-weight capability sandbox\footnote{FreeBSD
  Capsicum with minimal extensions in our prototype} to encapsulate
every runtime task in the data retrieval system's processing
pipeline. The sandbox is configured to allow all accesses that have
been validated by the OA without any further interception.  Second, a
\sys\ reference monitor (RM) runs in userspace, isolated within a
\textit{lwC}. It serves two purposes: It configures the sandbox when a
task starts and it validates any accesses that were not validated by
the OA ahead of time by making the required policy checks. In the
following, we describe these two components somewhat
abstractly. Section~\ref{sec:prototype} describes a concrete prototype
implementation of the RM and the sandbox on FreeBSD.

\if 0
While most \sys policy checks in a properly configured system should
be made by the OA ahead of time, \sys does rely on a limited amount of
runtime interception to make checks that cannot be made in
advance. The runtime interception consists of two components: A very
lightweight sandbox that is implemented in the kernel (with minimal
kernel changes), and a \sys reference monitor (RM), which runs in
userspace. Every runtime task in the pipeline is sandboxed; the
sandbox is configured to allow all accesses that have been validated
by the OA without any further interception. The RM serves two
purposes: It configures the sandbox when a task starts, and it
validates any accesses that were not validated by the OA ahead of time
by making the required policy checks. In the following, we describe
these two components somewhat abstractly. Section~\ref{sec:prototype}
describes a concrete prototype implementation of the RM and the
sandbox on FreeBSD.
\fi

\paragraph{Task registration}
When a new task starts, its access to all conduits is blocked by the
OS sandbox; the only thing the task can do is talk to the RM. To
get access to conduits, the task must register with the RM by
specifying which \emph{previously offline analyzed task} it
represents. For example, in \egsys, the task may register as the
indexer, the search engine or user X's worker for any known user X. The RM
records the choice and the taint provided during the OA for the
specified task.

Next, the RM looks up the last output of the OA for the specified task
to determine which accesses for the task have already been
validated. For each tuple (read, T, f, $conds$) or (write, T, f,
$conds$) in the output, the RM checks the state conditions $conds$,
and creates a list of all conduits and permissions for which the
conditions hold. It gives this list to the OS sandbox, which
subsequently allows the task these accesses directly.

For reasons of efficiency, our prototype implements the checking of
state conditions differently. There, the RM \emph{always maintains}
up-to-date lists of each task's valid accesses by tracking changes to
meta-data on which state conditions depend (e.g., friends lists of
region-specific blacklists in \egsys), and eagerly re-evaluating state
conditions when the meta-data changes. As a result, task registration
is very fast. The rationale for this implementation choice is
straightforward: In online systems like social networks, changes to
meta-data like friends lists are far less frequent than task
registrations (which happen once per user session), so tying the
expensive step of checking state conditions to meta-data changes
rather than task registrations results in less overhead.

Registering incorrectly, e.g., registering as the indexer in place of
Alice's worker or as Alice's worker in place of Bob's worker, either
maliciously or accidentally, cannot cause a policy violation in
\sys. However, doing so may cause expected accesses to be denied or
more accesses to fault into the RM thus slowing down the task.

\paragraph{Conduit access}
After a task has registered with the RM, it can open conduits for
reading and writing. Every conduit open call passes through the kernel
as usual. If the conduit and the mode (read/write) in which it is
being opened were provided to the OS sandbox as a valid access
during the task's registration, the kernel just allows the call. This
is the fast path and it should apply to most conduit accesses in a
properly configured system.

If, on the other hand, the OS sandbox does not know that the
specific access is valid, then it transfers control to the RM. The RM
then makes the same policy checks that the OA would have made for the
corresponding operation (read/write). The only difference is that the
RM does not generate any state conditions $conds$; it just checks them
immediately. If the checks succeed, the open call is allowed, else it
is denied.

\paragraph{Meta-data changes after a task registers}
As explained above, the OS sandbox is informed of a task's
pre-validated accesses when the task registers. A relevant question is
what to do when a subsequent meta-data change invalidates some of
these accesses. There are two choices here: Either the invalidated
accesses can be revoked in the OS sandbox or they can be left as
is. \sys chooses the latter option since revoking a permission from
the sandbox is costly.

This option is also secure since any access that the task does after
the invalidation could also have been done before the invalidation to
the same effect. An exception to this argument occurs when read access
is to be invalidated before the conduit is updated. In this case,
continuing the read access will allow the task to obtain the conduit's
updated content which it could not have obtained had the read access
been revoked immediately. To avoid such cases (when they are really a
concern), the system should be configured to store updated content in
new conduits (e.g., by versioning files). This is consistent with
systems like online social networks where existing content is updated
relatively infrequently (although fresh content is added quite
regularly).

\paragraph{Increasing task taints}
In most cases, a task's runtime taint is fixed when the task registers
and remains the same throughout its execution. In some cases, however,
the task may wish to increase its taint (i.e., make it more
restrictive) during its execution. For example, this is necessary if
the task wishes to read sensitive content after writing to a public
conduit. In this case, the task must start with a public taint and
acquire the taint of the sensitive content afterward.

\sys allows a task to increase its taint at runtime as
follows.\footnote{On the other hand, reducing a task's taint at
  runtime is not safe as this allows the task to leak previously read
  information. Therefore, \sys and all other coarse-grained taint
  tracking systems disallow reducing a task's taint.} At any point, a
running task may re-register as a new offline analyzed task whose
taint is higher than the task's current taint. In addition to making
all the checks that would be made during task registration, \sys also
checks that the policies of any conduits to which the task has open
write handles are more restrictive than the task's new taint. This
check is necessary to prevent leaks of data that the task reads under
the new taint. If this check fails, the re-registration is disallowed.

\paragraph{The overall cost of runtime interception}
The overall cost of runtime interception in \sys is generally very
low.  For interactive pipelines such as \egsys's RM interception
happens only a few times per user session (not per user query). For
instance, in \egsys, only four RM interceptions are needed per
session: (a) To register the worker task that serves the session (this
interception also validates the policies on the pipe that connects the
worker to the search engine), (b) When the worker accepts the user's
connection, (c) When the session is authenticated, and (d) At the end
of the session, to reset the worker task to a clean state for the next
user.

%% file: implementation.tex
\section{\sys\ prototype}
\label{sec:prototype}

Our {\sys} prototype runs on FreeBSD and relies on FreeBSD's kernel
capability support (Capsicum)~\cite{capsicum} and light-weight
contexts (\textit{lwC}s)~\cite{lwc} for sandboxing and isolation,
respectively. We briefly describe these primitives.

\paragraph*{Capsicum}
Capsicum is a OS sandbox to control a process' access to global
namespaces including the file system. Capsicum introduces a new mode
of process execution, the \emph{capability mode}. A process in this
mode can open new files (and, more broadly, make specific syscalls)
only if it has been granted the \emph{capabilities} to do so before it
entered the capability mode. {\sys} uses Capsicum directly to
implement its OS sandbox: Every task runs in capability mode with
capabilities to make only the accesses that were certified by the OA.

\paragraph*{Light-weight contexts (\textit{lwC}s)}
\textit{lwC}s are an OS abstraction that allow multiple tasks with
separate address spaces and separate file descriptor tables to
co-exist in the same process. \textit{lwC}s are orthogonal to
execution threads; a thread can switch between \textit{lwC}s. A
\textit{lwC} switch is more efficient than a process switch since the
former has no scheduler delays. 

{\sys} maps tasks one-one to \textit{lwC}s. The RM also runs inside a
privileged \textit{lwC}. The kernel is configured to redirect any
syscall outside a task's capability set to the RM \textit{lwC}. As
compared to a design that uses processes for the same purposes, this
design allows for faster switching between tasks and the RM, and for
faster resetting of tainted worker tasks at the end of user sessions
by avoiding scheduler delays.

\paragraph{Application life cycle}
An application is loaded with a customized script that first
initializes the RM \textit{lwC} in each process. Next, it initializes
application tasks in separate \textit{lwC}s and confining them with
Capsicum's capability mode. Then, the RM is invoked to register each
application task, giving it the capabilities to access anything that
was already certified by the OA and whose state conditions
hold. Depending on the type of a conduit, the capability to access it
takes different forms:
\begin{itemize}
\item[-] Files: The task is given \emph{Capsicum} capabilities to a
  small set of directories that contain hard links to all files that
  should be accessible to the task. (These directories and the hard
  links are created offline at the end of the OA and kept up-to-date
  by the RM as state conditions change.)
\item[-] Key-value (KV) tuples: For these, the RM relies on KV
  filters. The RM opens a socket to the KV store and installs a KV
  filter that limits access to only those tuples that are accessible
  to the task. It then gives this open socket to the task.
\end{itemize}
During its execution, an application task makes most conduit accesses
directly using the capabilities described above. For the few accesses
that are beyond these capabilities, it faults into the RM, which makes
policy checks.

\if 0
\paragraph{Kernel modification}
To support the {\egsys} application and other similar search-based
pipelines, we made a small modification to the FreeBSD kernel. We
added a special class of pipes on which only open file descriptors,
but no data, can be transferred. Such a pipe is used by the search
engine task to return file descriptors for documents matching the user
query to the front-end's worker task. The point of preventing data
transfers on the pipe is to ensure that the search engine does not
accidentally send private data to the worker. Further, to ensure that
the search engine does not send a descriptor of Alice's private file
to a worker task connected to Bob, we modify the kernel to check that
when a file descriptor is transferred, the receiving task has the
capability to access the underlying file.

(Note that Thoth implements the link from the search engine to the
worker differently. In Thoth, the search engine sends an ascii list of
document ids matching a query to the worker over a regular pipe; the
worker then opens these documents itself. However, checking that the
search engine does not send anything else requires RM interception on
\emph{every write to the pipe}, to make sure that only ascii document
ids are being written. In contrast, \sys's method requires a RM check
only once per session to check that the pipe is of this special
class.)
\fi

\paragraph{Capsicum modifications}
To support {\egsys} and other similar search-based pipelines, we made
two modifications to Capsicum. First, we modified Capsicum to allow a
pipe without read and write permissions to be used to transfer open
file descriptors but not data. In {\egsys}, such a pipe is used by the
search engine to return file descriptors for documents matching the
user query to the front-end's worker task. Since data transfers on the
pipe are forbidden, even a buggy search engine cannot accidentally
send private data to the worker.

Second, we modified Capsicum to allow a task in capability mode to
transfer file descriptors to another task in capability mode only if
the receiving task already has access capabilities on all conduits
referenced by the file descriptors. With this feature in place,
Capsicum prevents a buggy search engine from transferring a descriptor
for Bob's private file to a front-end worker task connected to
Alice. To implement this feature, we modified Capsicum to maintain
every task's capabilities in a binary lookup tree. When a file
descriptor is transferred to a task, Capsicum looks up the binary tree
for a capability to the conduit referenced by the descriptor. This
lookup's complexity is logarithmic in the number of distinct
capabilities the task has. In {\egsys}, only the front-end tasks
receive file descriptors and these tasks have very few capabilities
(at most 5), so the lookup is very fast.

%% file: evaluation.tex
\section{Evaluation}
\label{sec:eval}

\newcommand{\baseline}{\textsc{\textbf{B}aseline}\xspace}
\newcommand{\dynamic}{\textsc{\textbf{D}ynamic}\xspace}

In this section, we present results of an experimental evaluation of
our \sys\ prototype. In particular, we measure the overhead of policy
enforcement in the data retrieval system {\egsys} described
in~\ref{sec:design:example}. We instantiated this system with the
widely-used Apache Lucene as the search engine~\cite{lucene}. All
experiments were performed on Dell R410 servers, each with 2 Intel
Xeon X5650 2.66 GHz 6 core CPUs, 48GB main memory, running FreeBSD
11.0 (x86-64) with support for light-weight contexts
(\textit{lwC})~\cite{lwc} and Lucene version 4.7 with a minor
modification to allow it to run in Capsicum's capability mode.%
\footnote{In \sys, all tasks, including the Lucene indexer, run in
  capability mode. In this mode, the syscall \textsc{open} is
  disallowed by Capsicum, so we had to modify Lucene to use
  \textsc{openat} instead. We found that \textsc{openat} is faster
  than \textsc{open} so, to remain fair, we use the modified Lucene in
  the two other configurations that we compare against as well.}
The prototype uses OpenSSL v1.0.2h.  The servers are connected to
Cisco Nexus 7018 switches with 1Gbit Ethernet links, which offer
enough network bandwidth for all our experiments. Each server has a
1TB Seagate ST31000424SS disk formatted under UFS, which contains the
OS installation and a 258GB static snapshot of English language
Wikipedia articles from 2008~\cite{wikiss}.

\textbf{Experimental setup.} In the following experiments, we compare 
the performance of \sys\ to two systems; \textit{(i)} a system 
that does not enforce policies (\baseline), and \textit{(ii)} a system 
that enforces policies via \textit{pure} dynamic analysis (\dynamic). 
We give more details about \dynamic in the following.

In \dynamic, each task has a \textit{current} taint, which represents
the combined policies of all the conduits the task has read.  A task's
taint can become more restrictive as the task reads more conduits.  To
enforce policy, a task's writes must \textit{(i)} satisfy the
\textbf{update} rule of the destination conduit's policy, and
\textit{(ii)} satisfy the declassification conditions of the task's
current taint. \dynamic is very similar to Thoth; for fair comparison,
our \dynamic implementation, like \sys, takes advantage of
\textit{lwCs} and Capsicum for efficient in-process isolation and
sandboxing. This yields better performance than the original Thoth
prototype, which isolates each user session and the reference monitor
in a separate process.

A process in \dynamic, like \sys, can have multiple Capsicum-sandboxed
user \textit{lwCs} (each terminates a user connection), and a
privileged monitor \textit{lwC}. Conceptually, the RM intercepts all
conduit open calls and writes to perform taint tracking and dynamic 
policy checks. As in Thoth, we optimize taint tracking by not invoking 
the RM during open calls and instead logging such calls in the kernel.
During a write, the RM is invoked, it checks the open call trace to 
update the task taints and then performs the policy check for the 
write.  To summarize, \dynamic and \sys are identical architecturally: 
A process has multiple Capsicum-sandboxed user \textit{lwCs} and a 
privileged monitor \textit{lwC}. Both systems also enforce the same 
policies. However, unlike \sys, which pushes most policy evaluation 
overhead to the offline analysis, \dynamic performs pure dynamic IFC: 
the underlying kernel intercepts I/O and directs it to the reference
monitor, which in turn tracks taint and performs policy evaluation at
runtime.

\subsection{Search throughput}
First, we measure \sys's overhead on search throughput. We drive the
experiment with the following workload.  We simulate a population of
40,000 users. Each user is assigned a friend list consisting of 100
randomly chosen other users, subject to the constraint that the
friendship relationship is symmetric. Each document in the Wikipedia
corpus is assigned either a public, private, or friends-only policy in
the proportion 50/30/20\%, respectively.  Private and friends-only
documents are assigned to a user picked uniformly at random from the
population. A total of 1.1\% of the corpus is censored in some
region. A censored document's policy allows declassification to an
external user only if the destination's blacklist file does not
blacklist the document.

In this experiment, 24 concurrent users issue queries in parallel. We
use query strings based on the popularity of Wikipedia page accesses
during one hour on April 1, 2012~\cite{wikiimages}. Specifically, we
search for the titles of the top 20K visited articles and assign each
of the queries randomly to one of the users. User sessions run for
lengths 1, 2, 4, 8, 16, or 32 queries. Additionally, we report the
throughput when users maintain their sessions for the duration of the
experiment (20k queries).

In our setup, two server machines execute a Lucene instance with 
different index shards. The front-end submits a search request to one 
Lucene instance, which in turn forwards the request to the other 
instance and merges the results from both shards. To maximize the 
performance of the baseline and fully expose the policy enforcement 
overheads, the index shards and parts of the corpus relevant to our 
query stream are pre-loaded into the servers' main memory caches, 
resulting in a CPU-bound workload. To ensure load balance, we 
partitioned the index into two shards of 22GB and 33GB, chosen to 
achieve approximately equal query throughput.

Table~\ref{tab:qthr} shows the average throughput over 40 runs of 20K
queries each, for \baseline, \dynamic, and \sys. The standard
deviation over the 40 runs was below 0.87\% across all configurations.

\begin{table}[!htbp]
\centering
\begin{tabular}{c|c|cc|cc}
\multicolumn{1}{c|}{ }  & \baseline & 
\multicolumn{2}{c|}{\dynamic} & \multicolumn{2}{c}{\sys} \\
Q/S & Avg. & Avg. & \footnotesize{overhead} & Avg. & 
\footnotesize{overhead} \\
\midrule
1		& 309.36 & 287.04 & \footnotesize{7.21\%} & 294.70 & 
\footnotesize{4.74\%} \\
2		& 313.90 & 298.29 & \footnotesize{4.97\%} & 305.60 & 
\footnotesize{2.64\%} \\
4		& 316.49 & 304.58 & \footnotesize{3.76\%} & 312.54 & 
\footnotesize{1.25\%} \\
8		& 317.60 & 308.07 & \footnotesize{3.00\%} & 316.17 & 
\footnotesize{0.45\%} \\
16		& 318.66 & 309.27 & \footnotesize{2.95\%} & 318.34 & 
\footnotesize{0.10\%} \\
32		& 318.95 & 310.11 & \footnotesize{2.77\%} & 318.64 & 
\footnotesize{0.10\%} \\
20k		& 319.13 & 311.10 & \footnotesize{2.52\%} & 319.08 & 
\footnotesize{0.02\%} \\
\bottomrule
\end{tabular}
\caption{Average search throughput in queries per second. Standard 
deviation was less than 0.87\% from the average in all cases. First 
column indicates the session length (queries per session -- Q/S).}
\label{tab:qthr}
\end{table}

The key result is that \sys's static analysis reduces the runtime
enforcement overhead to near zero for sufficiently long session
lengths (0.1\% at 16 queries and 0.02\% at 20k queries). The \dynamic
system, which relies on pure runtime enforcement but is otherwise
equivalent, has a runtime overhead of approximately 2.5\% for large
session lengths.\footnote{A 2.5\% overhead may seem small; but
  increasing the peak capacity of a large datacenter by 2.5\% to
  account for it has a substantial cost!} Even for short session
lengths, \sys's runtime overhead is substantially lower than
\dynamic's.

In \dynamic and \sys, the front-end creates a new \textit{lwC} for
every incoming user session. In \sys, the monitor \textit{lwC}
additionally performs the required runtime checks associated with the
connected user's taint before granting access capabilities.  The
overheads of setting up new sessions (creating \textit{lwCs} and
performing runtime checks) dominate the policy enforcement overhead
for short sessions. At one query per session, \dynamic incurs a 7.21\%
overhead, whereas \sys incurs 4.74\%. Here, \sys outperforms \dynamic
since \textit{(i)} \sys performs fewer (runtime) checks compared to
\dynamic's full policy evaluation for all documents accessed per
query, and \textit{(ii)} \dynamic tracks the search engine's taint and
intercepts its writes to evaluate them against the search engine's
current taint, whereas the search engine's accesses within its
capability set are not intercepted in \sys. Furthermore, as the
session length increases, the cost of \sys's per-session setup costs
and runtime checks amortize over the session's queries, whereas
\dynamic performs full policy evaluation for each query.

At session length 20k, \sys incurs 0.02\% overhead, significantly
better than \dynamic's 2.52\% overhead. \sys's remaining runtime
overhead is due to the kernel's capability checks; in particular, when
the search engine attempts to send file descriptors corresponding to
the search results, the kernel checks that the front-end has existing
access capabilities for these descriptors.  This check is efficient;
its runtime complexity is logarithmic in the number of distinct
(directory) capabilities the receiving front-end has. In our
prototype, a front-end that satisfies runtime checks acquires few
directory capabilities\footnote{One on the connected user's hard links
  directory, another on named pipes to submit queries to the search
  engine, and three on directories with public documents.},
making the check light-weight.

\begin{figure}[tb] \centering
\epsfig{figure=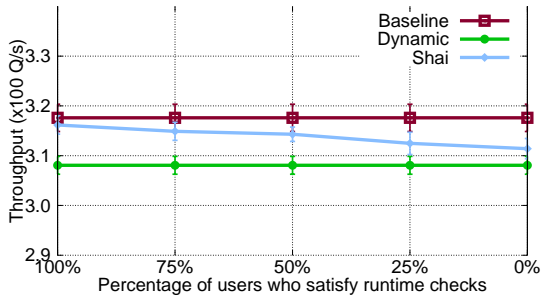, height=4cm,angle=0}
\caption{Average search throughput in queries per second of 24 
concurrent users, with sessions of length 8 queries. We included 
\baseline and \dynamic performance (upper and lower lines, 
respectively) for reference. Error bars show standard deviation.}
\label{fig:search_dep}
\end{figure}

In the previous experiment, all users connected from the regions that
were anticipated during the offline analysis. To quantify the
overhead of runtime checks required when runtime conditions deviate
from those expected, we next perform an experiment in which we vary
the proportion of users who connect from regions different from those
assumed during offline analysis.  Figure~\ref{fig:search_dep}
shows the average throughput over 40 runs of 20K queries each. The
error bars indicate the standard deviation over the 40 runs, which was
less than 0.72\% in all cases. We report the average throughput for
sessions of length 8 queries, but the following conclusions regarding
the relative overheads of \sys and \dynamic hold across all session
lengths.

With 100\% of users connecting from the expected region (i.e., all
user sessions satisfy the runtime checks associated with their taint),
\sys performs 316.17 Q/s (0.45\% overhead over \baseline), as in
Table~\ref{tab:qthr}. As the proportion of users who connect from
their expected regions decreases, \sys's performance declines
approximately linearly and approaches that of \dynamic, but never gets
worse. Even when all users connect from unexpected regions, \dynamic
incurs more overhead than \sys because it intercepts the search
engine's writes to evaluate them against the search engine's current
taint.  \sys's throughput degrades similarly if other runtime variables
are mispredicted. For instance, it declines approximately linearly
with the proportion of policies changed and the proportion of new
content added since the last offline analysis (not shown due to space
constraints).  This result shows that \sys's benefits decline
gracefully with the accuracy and freshness of the offline analysis.

\subsection{Scaling search throughput}
\label{sec:eval:scaling}

The throughput of a single Lucene search engine is relatively modest,
which raises the question of how much overhead \sys might impose on a
much faster system.  In the next set of experiments, we study \sys's
overhead in a replicated search engine configuration, and in a
configuration with a hypothetical search engine that has much higher
throughput than Lucene.

\textbf{Replication.} We performed the throughput experiment on a
replicated setup. In this experiment, four server machines execute
Lucene instances, where each index shard is replicated on two
servers. A front-end submits a search request to a lightly loaded
Lucene instance, which in turn forwards the request to another lightly
loaded instance processing the other shard and merges the results from
both shards. Here, 48 users issue queries in parallel, users maintain
their sessions for the duration of the experiment, and we measured the
average throughput over 40 runs, each 20K queries. \baseline,
\dynamic, and \sys all achieved an average throughput of almost
exactly twice (within 0.152\%) the respective throughput reported in
Table~\ref{tab:qthr} at session length 20k. This shows that \sys
(like \dynamic) scales linearly as the search engine is replicated.

\newcommand{\fsetup}{\textsc{\textbf{S}etup$_{3K}$}\xspace}
\newcommand{\vfsetup}{\textsc{\textbf{S}etup$_{30K}$}\xspace}

\textbf{Hypothetical fast search.}  To study \sys's overhead in a
hypothetical data retrieval systems that serve tens of thousands of
search requests per second, we replaced the Lucence search engine with
one that picks results randomly from the set of documents accessible
by the user who issues the query.  We measure \sys's overhead over
\textit{(a)} a dummy search engine that performs over 3K Q/s
(\fsetup), and \textit{(b)} a dummy search engine that performs over
30K Q/s (\vfsetup). The dummy search engine busy waits to consume a
fixed number of CPU cycles in \fsetup before returning the search
results, whereas it returns the results immediately without busy
waiting in \vfsetup. Note that \vfsetup represents an extreme
situation, shown here only to fully expose \sys's overheads; we do not
expect any realistic search engine to attain such high per-node
throughput.

\begin{figure}[t] \centering	
\subfloat[\fsetup]{%
\epsfig{figure=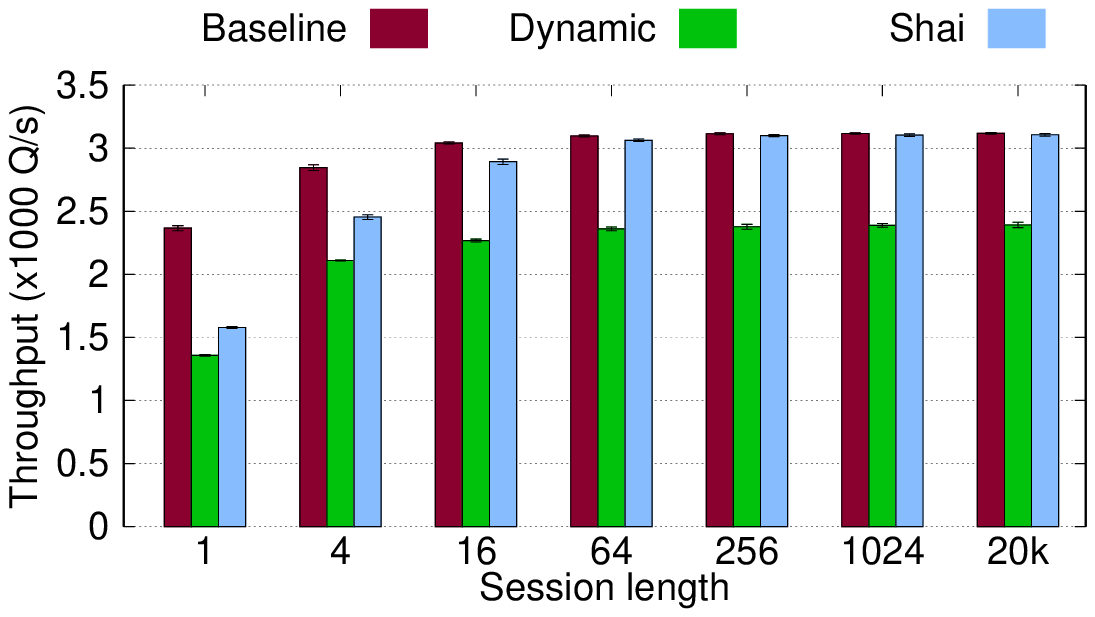, width=8.5cm,angle=0}%
\label{fig:scale:a}}
\vspace{-0.2cm}
\subfloat[\vfsetup]{%
\epsfig{figure=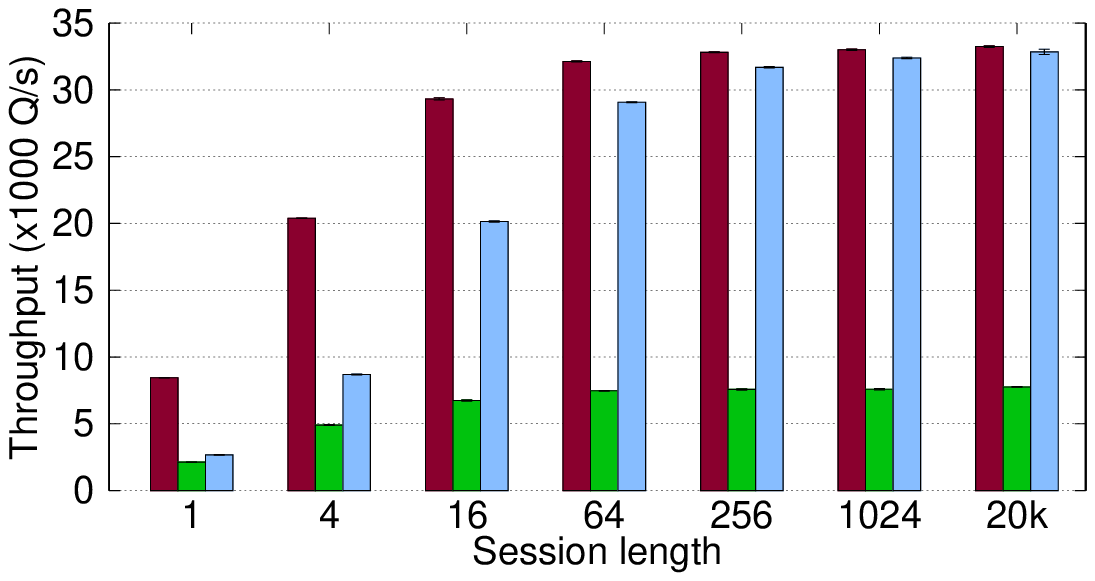, width=8.5cm,angle=0}%
\label{fig:scale:b}%
}
\caption{Search throughput in (Q/s) of 56 concurrent users, at 
different session lengths. Error bars show standard deviation.}
\label{fig:scale}
\end{figure}

In this experiment, a total of 56 concurrent users issue queries in
parallel to two server machines running the dummy search engine.  User
sessions run for lengths 1, 4, 16, 64, 256, 1024, and 20k queries.
Figure~\ref{fig:scale} shows the average throughput at the different
session lengths for \fsetup and \vfsetup (Figures~\ref{fig:scale:a}
and~\ref{fig:scale:b}, respectively). We report the average throughput
over 10 runs, each of length 30 seconds. Error bars show the standard
deviation across the 10 runs, which was below 0.9\% in all cases.

At small session lengths, both \sys and \dynamic have high overheads
due to the cost of creating \textit{lwCs} to isolate user sessions. As
the session length increases, the cost of session creation amortizes
across queries in \sys; the overheads are only 0.37\% and 1.2\% at
session length 20k, in~\ref{fig:scale:a} and~\ref{fig:scale:b}
respectively. These overheads are due to checking, at a high rate,
that the front-ends have existing capabilities over the transferred
file descriptors. On the other hand, \dynamic does not scale beyond
2.39K and 7.76K Q/s in~\ref{fig:scale:a} and~\ref{fig:scale:b},
because intercepting I/O to perform policy evaluation at runtime
limits performance. This result shows that \sys can maintain low
overhead even in very high-performance data retrieval systems.

\subsection{Search latency}
We next measure \sys's overhead on query latency. For this experiment,
a user issues one query at a time and waits until it receives a result
before issuing another query.  User sessions run for lengths 1, 2, 4,
8, 16, 32 or 4k queries.

\newcommand{\fallback}{\textsc{\textbf{S}hai}$_{mispredict}$}

\begin{table}[!htbp]
\centering
\begin{tabular}{ccccc}
Q/S & \baseline &  \dynamic & \fallback & \sys \\                      
\midrule
1			& 36.013 & 38.071 & 37.314 & 36.345 \\
2			& 36.007 & 37.843 & 37.122 & 36.108 \\
4			& 36.005 & 37.742 & 37.084 & 36.037 \\
8			& 35.981 & 37.719 & 37.021 & 36.008 \\
16			& 35.939 & 37.657 & 36.916 & 35.955 \\
32			& 35.928 & 37.599 & 36.904 & 35.941 \\
4k	& 35.905 & 37.597 & 36.913 & 35.960 \\
\bottomrule
\end{tabular}
\caption{Average query latency (ms). Standard deviation was less than
  0.8\% in all cases. The first column indicates the session length
  (queries per session -- Q/S).}
\label{tab:qlat}
\end{table}

Table~\ref{tab:qlat} shows the average query latency across 5 runs of
4K queries. Since \sys's overhead relies on satisfying the runtime
checks necessary to acquire capabilities, we report \sys's performance
when \textit{(i)} the user logs in from a geographic location
different than the region used during offline analysis
(Table~\ref{tab:qlat} column 4: \fallback), and when \textit{(ii)} the
user logs in from the geographic location that is used in the offline
analysis (Table~\ref{tab:qlat} column 5).

\sys's policy enforcement overhead on query latency is very low (at 
most 0.34ms). Also, \fallback~(when the user fails to satisfy 
the runtime checks to acquire capabilities) achieves better performance 
than \dynamic. The performance difference is due to tracking the taint 
and intercepting the search engine writes in \dynamic, whereas the 
search engine's accesses within its capability set are not intercepted 
in \fallback.

\subsection{Offline analysis}
Next, we measure the cost of running the offline analysis
over the expected data flows within the data retrieval system. The
runtime of the analysis depends on the number of tasks, the number of
expected accesses and relevant policies of each task, and the number
of accesses certified (each certified access may require creating a
hard link in the task's capability directory). We limit
the analysis to a single CPU. The analysis can be sped up by using
more CPUs since its computation is embarrassingly parallel
(except when creating hard links within the same directory).

\paragraph{Indexer and search engine flows}
The analysis takes under 2 seconds to process the flows of the search
engine and the indexer tasks against the entire Wikipedia corpus
($\sim$14.5 million documents subject to $\sim$80K different
policies).  Here, the searchable documents' policies permit read to
the indexer and the search engine, and the analysis grants both a
single capability for the top level directory of searchable documents.

\paragraph{Users flows}
Next, we measure the analysis time and storage requirements for the
users of the data retrieval system. For this experiment, we assume a
fixed default geographic location for every user. For each user, the
offline analysis checks the front-end's accesses of all public
documents, the user's private documents, and the friends-only
documents of the user's friends.
%% The granted access
%% capabilities are subject to runtime checks to validate the user's
%% identity and the geographic location (s)he is connected from.

We ran the OA on accesses of 100 users picked at random from the
population. Processing the accesses took 90.5 seconds per user on
average. This can be optimized using a faster storage medium since
most of this time (96.1\%) was spent waiting for the magnetic disk to
record hard links for access capabilities. To quantify potential
speed-up when using a ramdisk, we ran the offline analysis for the
same 100 users on a Dell R640 server machine with 385GB main memory,
and limited the analysis to use only one core of its Xeon Gold 6142
2.60GHz CPUs. This server machine has enough memory to store the
entire Wikipedia corpus in ramdisk, allowing us to create hard links
in ramdisk too. Using the ramdisk to store hard links, processing each
user took under 1.2 seconds on average (20\% of which was spent
creating hard links).

Each user's access capabilities consumed 12.9MB of disk space to store
145.8K hard links on average. Tasks' taints and state conditions
consumed less than 11MB of disk space for all 100 users combined.

\subsection{Indexing}
Finally, we measure the overhead of policy enforcement on the index
computation. We run the Lucene indexer over the entire 258GB snapshot
of the English Wikipedia. The resulting index is 54GB in size. Table 3
shows the average indexing time in minutes across 3 runs. The standard
deviation was less than 2\% in all cases.

\begin{table}[!htbp]
\centering
\begin{tabular}{l|cc}
 & Average	& Overhead	\\
\hline
\baseline	& 652.27 &			\\
\dynamic	& 672.02 & 3.02\%	\\
\sys		& 656.16 & 0.59\%	\\		
\end{tabular}
\caption{Indexing time in minutes.}
\label{tab:index}
\end{table}

\if 0
\begin{table}[!htbp]
\centering
\begin{tabular}{l|cc|rc}
\multicolumn{1}{r|}{Dataset}  & \multicolumn{2}{c|}{258GB} & 
\multicolumn{2}{c}{5GB} \\
 & Avg.	& Overhead	& Avg.	& Overhead	\\
\hline
\baseline	& 652.27 &			& 9.81 	& 			\\
\dynamic	& 672.02 & 3.02\%	& 10.08	& 2.74\%		\\
\sys		& 656.16 & 0.59\%	& 9.86 	& 0.56\%		\\		
\end{tabular}
\caption{Indexing time in minutes.}
\label{tab:index}
\end{table}
\fi

Enforcing policies with \sys\ during indexing incurs a runtime
overhead of 0.59\%, which is significantly lower than \dynamic's
3.02\%. \sys's overhead is due to the fact that the indexer creates
many new files, and all these file creations must be intercepted to 
ensure that output has appropriate policy given the indexer task's 
taint.  Policy enforcement in \dynamic additionally intercepts the 
indexer's writes to the index files and tracks the indexer's taint.

%% and intercepting writes to ensure that output has appropriate
%% policy given the indexer task's current taint.

Since indexing is a relatively infrequent operation in a search
pipeline, we believe that a runtime overhead of 0.59\% is
acceptable. However, in other systems where frequent file creation
occurs on the critical path, runtime interception of file creation
could be avoided as follows. Using an appropriate Capsicum capability,
we can restrict file creation to a specific directory with an
appropriate policy. All files created in this directory implicitly
inherit this policy. The offline analysis can check upfront that the
task creating the files can write to any file with this policy.

\subsection{Fault-injection tests}

To double-check \sys's ability to stop data leaks, we ran the fault
injection tests that were originally run on Thoth in~\cite{thoth}. In
all tests, \sys stopped all injected data leaks. (We refer the reader
to~\cite{thoth} for details of the tests.)

%% file: related.tex
\section{Related work}
\label{sec:related}

The work most closely related to \sys is obviously Thoth. We have
already described how \sys is a significantly more efficient re-design
of Thoth and how the two differ. In the following, we describe other
closely related work. 

\if 0
\thoth~\cite{thoth}, like \sys, is a policy compliance tool for data 
retrieval systems. In fact, the policy language used in \sys\ is 
inspired by {\thoth}'s policy language. (\sys additionally introduces 
policy predicates that refer to the underlying capability system.) 
Hence, both systems express similar policies. However, \thoth\ and 
\sys\ differ in enforcement techniques and architecture. \thoth\ uses 
pure dynamic analysis: \thoth intercepts I/O, tracks taint, and 
evaluates policies at runtime. Dynamic analysis incurs runtime overhead 
making it impractical for large-scale providers. On the other hand, 
\sys\ performs static flow analysis on the platform's dataflow graph, 
policies, and access control lists to predict the set of policies each 
data-handling process will be subject to at runtime. These policies are 
then enforced at runtime with a set of fine-grained I/O capabilities 
that are enforced directly by the OS, incurring very low runtime 
overhead suitable for providers running at a larger scale. 
Architecturally, \thoth\ relies on the underlying OS process isolation 
to isolate the reference monitor and users' sessions. In contrast, \sys 
relies on efficient OS isolation primitives to isolate the reference 
monitor and user sessions in the same process.
\fi

Grok~\cite{Sen2014sp} is a privacy compliance tool that is deployed on
the backend data analysis pipelines of the Bing search engine. As
opposed to \sys, which enforces data-specific policies, Grok enforces
only type-specific policies. As a result, Grok's analysis can be, and
is, entirely static; there is no runtime component and no runtime
overhead. Grok's analysis is meant to detect bugs and
misconfigurations, not strictly enforce policies. In fact, the
analysis uses possibly unsafe (but very scalable) heuristics and it
can have both false positives and false negatives, which must be
resolved manually. Nonetheless, Grok demonstrates that policy
enforcement can scale to actual production pipelines.

The idea of runtime coarse-grained taint tracking via kernel
interception was pioneered by the operating systems
HiStar~\cite{histar} and Asbestos~\cite{asbestos}, and later developed
in Flume~\cite{flume}. However, these operating systems assign
\emph{abstract} taints to processes; the mapping from taints to
policies, as well as the enforcement of declassification is left to
trusted processes. In contrast, Thoth and \sys enforce declarative
policies (that also include declassification conditions) directly.

\paragraph*{Hybrid policy enforcement}
There are a number of other systems that combine static (offline)
analysis with runtime monitoring for security. Such techniques exist
for information flow control
(IFC)~\cite{static4hifc,hybridstaticruntime,hlio}, enforcing safety
properties~\cite{p.enf.abs.ref,sasi}, and gradual
typing~\cite{gradualIFC,gradualtyping}. Fredrikson
et.~al~\cite{p.enf.abs.ref} use abstraction refinement and model
checking to instrument code with sufficient checks to enforce
policies, and Rocha et.~al~\cite{hybridstaticruntime} use code
analysis to inject policy checks in program code to enforce IFC and
declassification policies. Moore and Chong use static analysis to
reduce monitoring overhead by selectively marking variables which
cannot cause security violations to not be tracked at
runtime~\cite{static4hifc}. Similar to \sys, these approaches try to
perform as many checks as possible statically, and use runtime checks
only where static checks are impossible. However, all these systems
combine static and dynamic analysis at \emph{fine-granularity} and
require the source code of the application. In contrast, {\sys}'s
offline analysis (which can also be viewed as a static analysis) uses
only a description of the system pipeline, not the source or compiled
code of the system. Moreover, \sys combines static and dynamic
analysis at \emph{coarse-granularity}. As far as we know, {\sys} is
the first system to do this.

RIF is a policy model similar in concept to \sys/Thoth's policy
model. RIF has been implemented in an extension of the Java
programming language called JRIF~\cite{jrif}. Like the aforementioned
work, JRIF enforces policies at fine-granularity by hybrid analysis
consisting of mostly static inference and some runtime checks. All the
differences from \sys mentioned above apply to JRIF as well. JRIF's
declassification conditions are linked to specific program points, not
predicates on the system/conduit state as in \sys. It is unclear
whether a pipeline such as {\egsys} can be implemented in JRIF and, if
so, what the cost of the runtime checks would be.

%% In
%% contrast to all these systems that perform static analysis on
%% \textit{code}, \sys\ analyzes \textit{data flow graph} at process
%% granularity.  \sys's analysis is less permissive (since \sys\ cannot
%% map program inputs to outputs precisely), however, such analysis is
%% suitable for data retrieval systems where application codebases are
%% large and frequently updated, and often written in different
%% programming languages. Additionally, the integrity of \sys's runtime
%% monitor relies on the integrity of the underlying OS, rather than the
%% integrity of the programming language runtime. As far as we know,
%% \sys\ is the first system to track changes that affect the prior
%% decisions of the static flow analysis and to adjust them accordingly
%% as discussed in Section~\ref{sec:commitments}.

\paragraph*{Policy debugging}
A problem complementary to that addressed by {\sys} is that of
debugging policies. This problem has been addressed in prior work
using logic programming techniques~\cite{eon}, model
checking~\cite{difcmodel} and flow simulations~\cite{polsim}. Although
the problem is orthogonal to our goals, some of the techniques used
are similar. For example, PolSim~\cite{polsim} performs an analysis
similar to \sys's offline analysis, on the same policy
language. However, unlike \sys's goal of checking that accesses are
policy compliant, PolSim seeks to ensure that the entire pipeline
works, despite restrictions imposed by policies. Consequently, PolSim
outputs blocked data flows in the pipeline and suggestions for how to
change the policies to allow the flows.

%% file: discussion.tex
\section{Conclusion}
\label{sec:sum}

\sys\ shows that it is possible to enforce data-specific flow policies
in data retrieval systems with near-zero runtime overhead in the
common case. \sys\ relies on a combination of an offline flow
analysis, session-level binding of runtime variables, and light-weight
runtime monitoring using an OS capability sandbox to achieve this
goal. The key insight behind \sys\ is to push as much work as possible
to the offline analysis, often relying on anticipated values of
runtime parameters, and to use efficient OS techniques (light-weight
contexts and Capsicum capabilities) to minimize runtime overhead. This 
combination
keeps \sys's overheads very low, even when the system throughput is
very high.

%% file: shai.bbl
\begin{thebibliography}{10}

\bibitem{lucene}
{Apache Lucene}.
\newblock \url{http://lucene.apache.org}.

\bibitem{gdpr}
{The EU General Data Protection Regulation}.
\newblock \url{https://www.eugdpr.org/the-regulation.html}, January 2018.

\bibitem{polsim}
Mohamed Alzayat.
\newblock Polsim: Automatic policy validation via meta-data flow simulation.
\newblock Master's thesis, Saarland University, Saarbruecken, 2016.

\bibitem{hlio}
Pablo Buiras, Dimitrios Vytiniotis, and Alejandro Russo.
\newblock Hlio: Mixing static and dynamic typing for information-flow control
  in haskell.
\newblock In {\em Proceedings of the 20th ACM SIGPLAN International Conference
  on Functional Programming}, ICFP 2015, pages 289--301, New York, NY, USA,
  2015. ACM.

\bibitem{eon}
Avik Chaudhuri, Prasad Naldurg, Sriram~K. Rajamani, G.~Ramalingam, and
  Lakshmisubrahmanyam Velaga.
\newblock Eon: Modeling and analyzing dynamic access control systems with logic
  programs.
\newblock In {\em Proceedings of the 15th ACM Conference on Computer and
  Communications Security}, CCS '08, pages 381--390, New York, NY, USA, 2008.
  ACM.

\bibitem{asbestos}
Petros Efstathopoulos, Maxwell Krohn, Steve VanDeBogart, Cliff Frey, David
  Ziegler, Eddie Kohler, David Mazi\`{e}res, Frans Kaashoek, and Robert Morris.
\newblock Labels and event processes in the {A}sbestos operating system.
\newblock In {\em Proceedings of the 20th ACM Symposium on Operating Systems
  Principles (SOSP)}, 2005.

\bibitem{thoth}
Eslam Elnikety, Aastha Mehta, Anjo Vahldiek-Oberwagner, Deepak Garg, and Peter
  Druschel.
\newblock Thoth: Comprehensive policy compliance in data retrieval systems.
\newblock In {\em 25th USENIX Security Symposium (USENIX Security 16)}, 2016.

\bibitem{sasi}
\'{U}lfar Erlingsson and Fred~B. Schneider.
\newblock Sasi enforcement of security policies: A retrospective.
\newblock In {\em Proceedings of the 1999 Workshop on New Security Paradigms},
  NSPW '99, pages 87--95, 2000.

\bibitem{gradualIFC}
L.~Fennell and P.~Thiemann.
\newblock Gradual security typing with references.
\newblock In {\em 2013 IEEE 26th Computer Security Foundations Symposium},
  pages 224--239, June 2013.

\bibitem{p.enf.abs.ref}
Matthew Fredrikson, Richard Joiner, Somesh Jha, Thomas Reps, Phillip Porras,
  Hassen Sa\"{\i}di, and Vinod Yegneswaran.
\newblock Efficient runtime policy enforcement using counterexample-guided
  abstraction refinement.
\newblock In {\em Proceedings of the 24th International Conference on Computer
  Aided Verification}, CAV'12, pages 548--563, 2012.

\bibitem{jrif}
Elisavet Kozyri, Owen Arden, Andrew~C. Myers, and Fred~B. Schneider.
\newblock {JRIF}: Reactive information flow control for {Java}.
\newblock Technical report, Cornell University, 2016.

\bibitem{flume}
Maxwell Krohn, Alexander Yip, Micah Brodsky, Natan Cliffer, M.~Frans Kaashoek,
  Eddie Kohler, and Robert Morris.
\newblock Information flow control for standard {OS} abstractions.
\newblock In {\em Proceedings of 21st ACM SIGOPS Symposium on Operating Systems
  Principles (SOSP)}, 2007.

\bibitem{lwc}
James Litton, Anjo Vahldiek-Oberwagner, Eslam Elnikety, Deepak Garg, Bobby
  Bhattacharjee, and Peter Druschel.
\newblock Light-weight contexts: An os abstraction for safety and performance.
\newblock In {\em USENIX Symposium on Operating Systems Design and
  Implementation (OSDI 16)}, 2016.

\bibitem{static4hifc}
S.~Moore and S.~Chong.
\newblock Static analysis for efficient hybrid information-flow control.
\newblock In {\em 2011 IEEE 24th Computer Security Foundations Symposium},
  pages 146--160, June 2011.

\bibitem{hybridstaticruntime}
B.~P.~S. Rocha, M.~Conti, S.~Etalle, and B.~Crispo.
\newblock Hybrid static-runtime information flow and declassification
  enforcement.
\newblock {\em IEEE Transactions on Information Forensics and Security},
  8(8):1294--1305, Aug 2013.

\bibitem{Sen2014sp}
Shayak Sen, Saikat Guha, Anupam Datta, Sriram~K. Rajamani, Janice Tsai, and
  Jeannette~M. Wing.
\newblock Bootstrapping privacy compliance in big data systems.
\newblock In {\em Proceedings of the 35th IEEE Symposium on Security and
  Privacy (S\&P)}, 2014.

\bibitem{gradualtyping}
Jeremy~G. Siek and Walid Taha.
\newblock Gradual typing for functional languages.
\newblock In {\em IN SCHEME AND FUNCTIONAL PROGRAMMING WORKSHOP}, pages 81--92,
  2006.

\bibitem{capsicum}
Robert N.~M. Watson, Jonathan Anderson, Ben Laurie, and Kris Kennaway.
\newblock A taste of {C}apsicum: Practical capabilities for unix.
\newblock {\em Commununications of the ACM}, 55(3), March 2012.

\bibitem{wikiimages}
{Wikimedia Foundation}.
\newblock { Image Dump}.
\newblock \url{http://archive.org/details/wikimedia-image-dump-2005-11}.

\bibitem{wikiss}
{Wikimedia Foundation}.
\newblock Static {HTML} dump.
\newblock \url{http://dumps.wikimedia.org/}.

\bibitem{histar}
Nickolai Zeldovich, Silas Boyd-Wickizer, Eddie Kohler, and David Mazi\`{e}res.
\newblock Making information flow explicit in {HiStar}.
\newblock In {\em Proceedings of the 7th USENIX Symposium on Operating Systems
  Design and Implementation (OSDI)}, 2006.

\bibitem{difcmodel}
Mingyi Zhao and Peng Liu.
\newblock Modeling and checking the security of difc system configurations.
\newblock In Ehab Al-Shaer, Xinming Ou, and Geoffrey Xie, editors, {\em
  Automated Security Management}, pages 21--38. Springer International
  Publishing, 2013.

\end{thebibliography}
